\documentclass[manuscript]{acmart}
\AtBeginDocument{%
  }

\setcopyright{acmlicensed}
\copyrightyear{2026}
\acmYear{2026}
\acmDOI{}
\acmConference[XAIxArts 2026]{Explainable AI for the Arts Workshop 2026}{July 13, 2026}{London, UK}
\acmISBN{978-1-4503-XXXX-X/2018/06}

\begin{document}

\title{Exploring Normativity in Stable Diffusion: Insights for XAI in the Arts}

\author{Michelle Dutoit}
\email{michelle.dutoit@ifi.lmu.de}
\affiliation{%
  \institution{LMU Munich}
  \city{Munich}
  \country{Germany}
}
\affiliation{%
  \institution{ISIR, Sorbonne Universit\'e, CNRS}
  \city{Paris}
  \country{France}
}

\author{Baptiste Caramiaux}
\email{caramiaux@isir.upmc.fr}
\affiliation{%
  \institution{ISIR, Sorbonne Universit\'e, CNRS}
  \city{Paris}
  \country{France}
}

\renewcommand{\shortauthors}{Dutoit and Caramiaux}

\begin{abstract}
Generative text-to-image (T2I) systems are increasingly adopted in creative practice, yet their normative behaviors remain underexplored from the perspective of creative practitioners. In this workshop paper, we present a within-subject study with 14 creative practitioners using Stable Diffusion to create illustration from two tasks of differing specificity. We investigate whether and how practitioners perceive normative behavior in a T2I system and how it impacts their creative process depending on task specificity. Our findings show that participants perceived normative behavior through invariant patterns, stereotypical output, and unsolicited omission or addition of details. These experiences led to feelings of disempowerment and creative compromise. We discuss implications for XAIxArts, including prompt transparency and artist empowerment in creative contexts. 
\end{abstract}


\maketitle

\section{A Brief Background and Motivation}
T2I systems have been shown to misrepresent cultures and reinforce Western-centric norms~\cite{agarwal2025ai,donnarumma_against_2022}. As an example, these generative T2I systems inadequately represent South Asian culture by failing to generate culturally specific subjects, amplifying hegemonic defaults, and perpetuating stereotypical tropes~\cite{qadri_ais_2023}. Ethnographic work in Bangladesh further illustrates how such systems can limit creative ideation and yield distorted or inaccurate representations for local artists~\cite{mim2024between}. This relates to normativity of generative AI (genAI) systems, which have been shown to tend to replicate normative narratives, and rarely producing non-normative alternatives when not mentioned explicitly~\cite{gillespie2024generative}. We believe that XAIxArts could gain from an understanding of normativity. First, because the normative basis of the genAI tools leads to outputs that may not be understandable or transparent to the people using it. Understanding normativity and finding solutions through XAIxArts could help reduce its negative impacts and alleviate its influence. Second, studying how people perceive and deal with normativity progresses XAIxArts interventions. There has already been research on strategies for circumventing normative behaviors that have been shown to include repeated prompting, finding workarounds, iteratively refining prompts to achieve their objectives, and testing multilingual prompts \cite{taylor_-straightening_2025, qadri_ai_2025}. Further studying these strategies can provide insight on designing XAIxArts strategies.

Previous studies in the normative behavior of AI-based systems tend to illustrate the values and culture conveyed by these systems, and how users try to circumvent this gaze, but less so the processes through which these normative behaviors operate. 
This gap is particularly significant in creative and artistic contexts, where T2I systems are increasingly adopted for ideation and production~\cite{rajcic2024towards}. Artists develop strategies to achieve the desired results, such as creating prompt templates and adapting their language to match their expectations~\cite{chang_prompt_2023, goloujeh2024promptjourney}. Recent research also reported shortcomings in the use of T2I in creative tasks, including the difficulty in expressing thoughts in words \cite{mccormack_is_2023}, and ``concept entanglement'', which refers to the system's inability to represent specific concepts independently of other concepts~\cite{sanchez_examining_2023}. 

Building on this work, we examine the normative behavior of T2I systems (in particular a diffusion-based system) from the perspective of creative practitioners who are new to these tools. 
Our research question is ``\textbf{How do creative practitioners perceive the normativity of the T2I system depending on the task specificity, and how does the normative behavior affect them?}''. 

\section{Study Design}
 We recruited 14 visual creative practitioners (graphic designers, illustrators, UX designers with illustration background, and others) for a within-subject study located in Western Europe. The study was reviewed and approved by the institution's Research Ethics Committee. Participants were recruited through the authors' networks and calls advertized on social media. Our inclusion criteria were visual creative practitioners with no or minimal prior T2I experience, to avoid pre-formed strategies for circumventing T2I systems' limitations and normative constraints.
 
The study was conducted in person or online using a video conferencing tool. In the beginning, the participants received an information letter and completed a consent form. Then participants received two tasks,  each requiring them to create an illustration from a given source medium using the Stable Diffusion WebUI \cite{AUTOMATIC1111_Stable_Diffusion_Web_2022}. We chose to use Stable Diffusion, as local deployment gave us control over generation speed and removed prompt-usage restrictions. In the first task, the stimulus was a 30-second audio clip, evoking an outdoor hiking environment, featuring sounds of footsteps, birds, and wind. In the second task, participants received a short news article describing a man in his 40s and his grandmother in her 90s visiting all U.S. national parks together, which is disclosed in the Appendix \ref{app:story}. We chose this story because it depicts a grandmother-grandson relationship with both of older age, a pairing that is rarely represented in illustrations and, therefore, difficult to generate with genAI. Tasks were presented in a fixed order rather than counterbalanced to prevent the narrative details of the story from influencing participants' interpretation of the soundscape. This fixed order preserved exploratory freedom in the first task, though it may have introduced risks of order, practice, and fatigue effects.
Finally, each session concluded with a semi-structured interview, which addressed how the participants perceived the normativity of the system, specifically by asking how they evaluated expressive freedom, assistance provided by the system, and perceived neutrality of the tool.

We analyzed the interview data as follows: The recordings were transcribed and translated locally on a secure computer with Whisper from OpenAI \cite{radford2023robust}, as the interviews were conducted in English, German, and French. The interviews were edited for the paper in intelligent verbatim for readability and the translations were reread to ensure validity. Data were analyzed using inductive thematic analysis, as described in Braun and Clarke \cite{braun2012thematic, braun2019reflecting, braun2021one}. Both researchers familiarized themselves with the data, and independently created codes. They consolidated them together in a shared codebook, derived themes and sub-themes, and refined or eliminated them until a consensus was reached.

\section{Highlights from our Analysis}
Our findings indicate that participants identified normative behavior in the system through three main patterns: stereotypical imagery, invariant responses to prompts, and the omission or unsolicited addition of details.

First, participants identified the system's normative behavior as coming from stereotyped, biased imagery. One participant for example
observed that the system ``\textit{conveys the western culture norms, middle class}'' and has ``\textit{a commercial, low key visual culture.}''. 
This aligns with prior work documenting biases in generative models, including gender-profession associations~\cite{luccioni_stable_2024} and Western-centric depictions of non-Western cultures~\cite{qadri_ais_2023}. Some example images are shown in Figure \ref{fig:example_images}.

Participants also noticed repeated compositional patterns as another manifestation of the system's normative behavior. Several participants observed shared commonalities in layout, perspective, and framing, although they often struggled to articulate them precisely. One participant probed these invariants deliberately, repeating the same prompt words and inspecting what remained consistent across outputs.
Another way in which participants perceived normativity in the system's behavior was through details specified in the prompts, but ignored by the model. Participants noted this in the interviews, describing normativity through omitted details, particularly in the more specific task (Story Task). For example, the male character was frequently left out or depicted as the same age as the older woman, as well as ignored style cues such as layout, line trace, and color. They theorized that the system performed better with less specified prompts, as too many details in prompts were often ignored. However, participants observed that when fewer details were included, the system generated imagery that conformed more closely to normative representations, such as that the older woman was generated with white hair, glasses, and a cane. One participant turned this into a deliberate strategy, abandoning the grandmother-grandson concept entirely to probe what non-normative content the system could produce.

Beyond perceiving the norms, it impacted the behavior of the participants. It contributed to a sense of disempowerment, as captured by one participant: ``\textit{So you could say that I assisted the tool, but I don't know if the tool needed it.}''. A common response was to abandon efforts to influence aspects perceived as uncontrollable, raising the possibility that users may forgo non-normative creative goals when faced with the system's constraints. Participants resorted to various compromises on their initial concept and tried to find a satisfactory outcome within the norms. They adjusted their concepts to match perceived system capabilities, adopted conventional prompt language (e.g. prompting a family portrait), or abandoned their initial concept altogether in favor of iterating on whatever the system first produced.

\vspace{4mm}
        \begin{minipage}[t]{\linewidth}
            \centering
            \includegraphics[width=0.15\linewidth,alt={On the left image is an conventional illustration of a person hiking in nature, the sun is rising and there are mountains and birds.}]{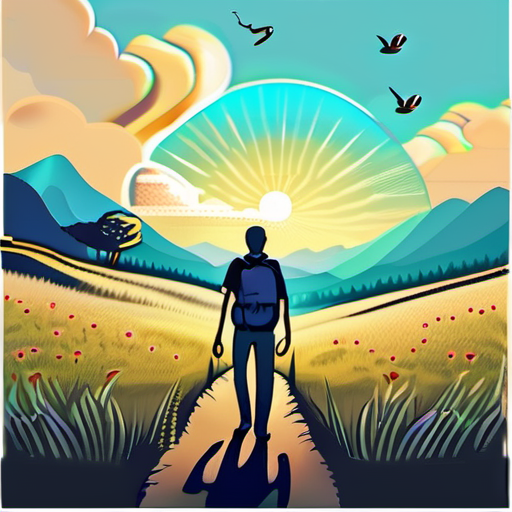} 
            \includegraphics[width=0.15\linewidth,alt={The second to left image shows another conventional illustration of cows grazing in the meadow next to them a tree, and in front of them a street.}]{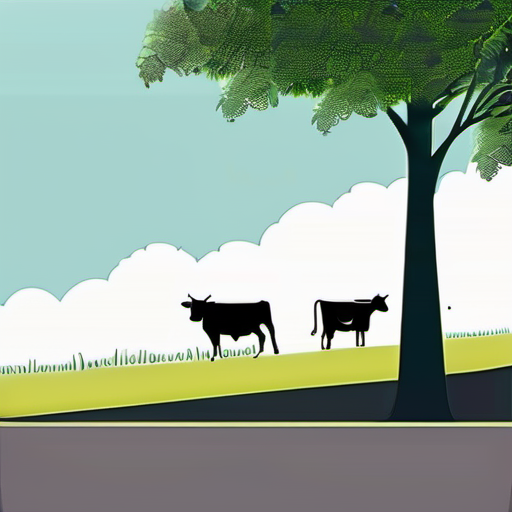} 
            \includegraphics[width=0.15\linewidth,alt={The image in the middle is a portrait of an older woman with white hair and a hat laughing with the sun in the background.}]{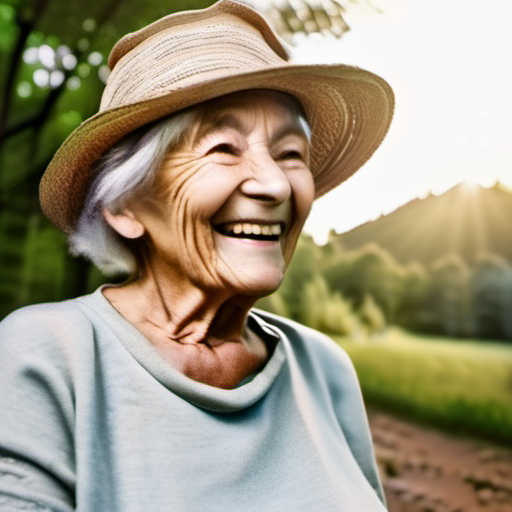}
            \includegraphics[width=0.15\linewidth,alt={The second to right image is a conventional illustration of an older couple holding hands and hiking through fields, with mountains in the background.}]{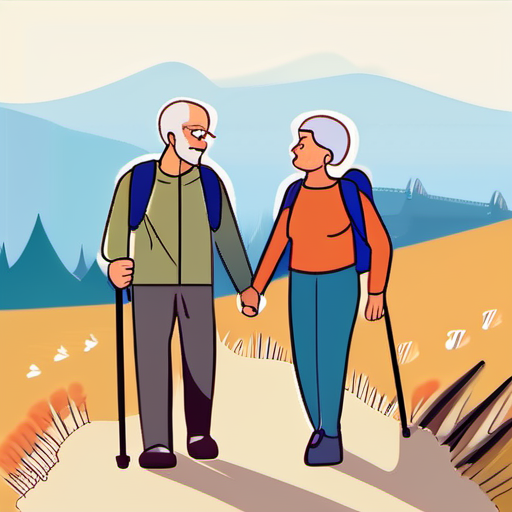}  
            \includegraphics[width=0.15\linewidth,alt={On the right most image is a illustrated portrait of an older woman with a young boy hugging, with a river and mountains in the background.}]{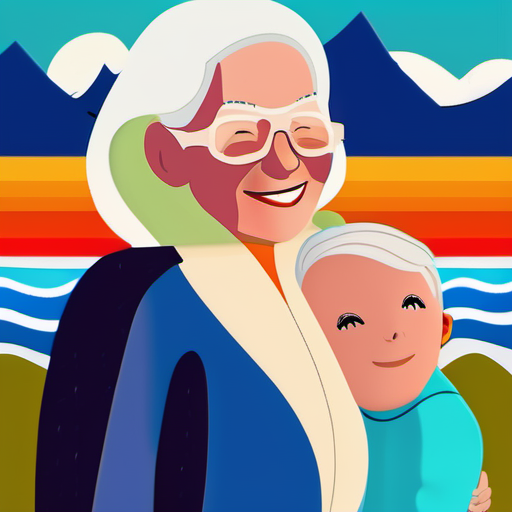} 
            \captionof{figure}{Examples of Generated Images }
            \label{fig:example_images}
        \end{minipage}

\section{Discussion}
This work aims to contribute to the development of XAIxArts in three ways. First, we highlight XAI as an opportunity to go beyond default behavior. Second, we present prompt transparency as a promising research direction in this field. Third, we discuss the feeling of disempowerment stemming from the normative behavior of T2I and the extent to which XAI could prevent it.

Participants perceived normative behavior, 
which contributed to a simplification of visual representation that favored default depictions. This is significant as users tend to adhere to default options, representing limited perspectives and restricted visibility of alternative viewpoints \cite{gillespie2024generative}, while artists use default abstractions and ‘make do’ with them, unable to explore other possibilities \cite{li_beyond_2023}. Advancing research on default behaviors is therefore essential, including exploring strategies for moving beyond them, through e.g., leveraging rather than averaging out stochasticity to circumvent inherent biases \cite{qadri_ai_2025}, and by enabling deeper interaction with training data through techniques such as LoRA \cite{hu2022lora}, or DreamBooth \cite{ruiz2023dreambooth}. 
XAI can offer strategies to overcome default behaviors, such as interaction focused approaches~\cite{liao2021human} or black-box explanation tools~\cite{ribeiro2016should,lundberg2017unified}. By allowing visibility and designing for glass-box abstractions \cite{smith2026noise}, users can gather more insight into the nature of diffusion and its capabilities as their experience grows \cite{shneiderman2007creativity}.

The system’s deviations from participants’ prompts, such as selectively removing or adding details without explanations, made participants reflect on the decisions made by the system and highlights the importance of prompt transparency. 
Such transparency aims to enable relevant stakeholders to develop an accurate understanding of the system’s capabilities, limitations, and mechanisms for controlling its outputs~\cite{liao2023ai}. While transparency has primarily been investigated in high-stakes application domains, we argue that it deserves greater attention in the creative domain, given the role creative practice plays in shaping our cultural fabric. In a recent study, artists also discussed the benefit of transparency in fostering creative practice while informing about the appropriate uses of T2I~\cite{shelby2024generative}.
Relevant to prompt transparency, existing techniques such as LIME~\cite{ribeiro2016should}, SHAP~\cite{lundberg2017unified}, and neural integrated gradients~\cite{sundararajan2017axiomatic} have been employed to enhance transparency in language models by highlighting significant tokens in the input that contribute to a given output. Extending these techniques to T2I models could help highlight omitted elements in prompts. However, a complementary approach is needed to explicitly identify and make transparent the details added by the system. One potential solution is the use of generated captions to highlight discrepancies between the input prompts and the outputs.
Researching prompt transparency could thus enable  more effective application and appropriation of T2I models as creative tools and cultural instruments.

Participants in our study frequently expressed frustration and a sense of powerlessness when using the system. Mahdavi Goloujeh et al. \cite{goloujeh2024promptjourney} similarly found that users with focused objectives and well-defined visions often experienced frustration when the system failed to generate important elements of their visions, linking this to a loss of user autonomy. This is an interesting contrast, as genAI systems are typically marketed, for example, Stability AI describes Stable Diffusion as ``empowering billions of people to create stunning art within seconds.'' \footnote{\url{https://stability.ai/news/stable-diffusion-announcement}}. Disempowerment through genAI remains underexplored. Previous research has documented how T2I output can disempower communities by perpetuating negative stereotypes and narratives of marginalized groups~\cite{10.1145/3613904.3642166, qadri_ais_2023}, and more generally, the feeling of disempowerment and loss of autonomy can produce a \textit{chilling effect} \cite{jiang2023ai}, and dissuade creatives from engaging with the development of creative-centered AI tools. The ``democratization of art''\footnotemark[1] is negated through the harms these technologies bring to the creative community.
Recent advances that offer users greater control, such as the fine-tuning methods discussed above, represent a promising direction. However, these approaches also introduce new challenges, including the extensive tinkering necessary to identify effective adjustments~{\cite{Cui-2025}}.  This feeling of disempowerment goes against an important objective of system design, that a tool should empower the user to allow them to achieve their goals. Li et al.~\cite{li_beyond_2023} argue that it is important to investigate how a tool influences the creative situation beyond the usability and functionality. Therefore, an important research challenge is to investigate more closely the construct of disempowerment in genAI-based creative tool practice. This can be done by furthering research on the disempowerment, which directly relates to XAIxArts, as a central theme is the empowerment of creatives. Working directly with artists helps to design interventions that can specifically address them and their empowerment, such as developing their AI literacy, designing accessible and inclusive XAI tools, and engaging them for input and feedback~\cite{bryan2025xaixarts}. 
\bibliographystyle{ACM-Reference-Format}
\bibliography{sample-base}


\begin{thebibliography}{33}


\ifx \showCODEN    \undefined \def \showCODEN     #1{\unskip}     \fi
\ifx \showISBNx    \undefined \def \showISBNx     #1{\unskip}     \fi
\ifx \showISBNxiii \undefined \def \showISBNxiii  #1{\unskip}     \fi
\ifx \showISSN     \undefined \def \showISSN      #1{\unskip}     \fi
\ifx \showLCCN     \undefined \def \showLCCN      #1{\unskip}     \fi
\ifx \shownote     \undefined \def \shownote      #1{#1}          \fi
\ifx \showarticletitle \undefined \def \showarticletitle #1{#1}   \fi
\ifx \showURL      \undefined \def \showURL       {\relax}        \fi
\providecommand\bibfield[2]{#2}
\providecommand\bibinfo[2]{#2}
\providecommand\natexlab[1]{#1}
\providecommand\showeprint[2][]{arXiv:#2}

\bibitem[Agarwal et~al\mbox{.}(2025)]%
        {agarwal2025ai}
\bibfield{author}{\bibinfo{person}{Dhruv Agarwal}, \bibinfo{person}{Mor Naaman}, {and} \bibinfo{person}{Aditya Vashistha}.} \bibinfo{year}{2025}\natexlab{}.
\newblock \showarticletitle{AI suggestions homogenize writing toward western styles and diminish cultural nuances}. In \bibinfo{booktitle}{\emph{Proceedings of the 2025 CHI Conference on Human Factors in Computing Systems}}. \bibinfo{pages}{1--21}.
\newblock


\bibitem[AUTOMATIC1111(2022)]%
        {AUTOMATIC1111_Stable_Diffusion_Web_2022}
\bibfield{author}{\bibinfo{person}{AUTOMATIC1111}.} \bibinfo{year}{2022}\natexlab{}.
\newblock \bibinfo{booktitle}{\emph{{Stable Diffusion Web UI}}}.
\newblock
\urldef\tempurl%
\url{https://github.com/AUTOMATIC1111/stable-diffusion-webui}
\showURL{%
\tempurl}


\bibitem[Braun and Clarke(2012)]%
        {braun2012thematic}
\bibfield{author}{\bibinfo{person}{Virginia Braun} {and} \bibinfo{person}{Victoria Clarke}.} \bibinfo{year}{2012}\natexlab{}.
\newblock \bibinfo{booktitle}{\emph{Thematic analysis.}}
\newblock \bibinfo{publisher}{American Psychological Association}.
\newblock


\bibitem[Braun and Clarke(2019)]%
        {braun2019reflecting}
\bibfield{author}{\bibinfo{person}{Virginia Braun} {and} \bibinfo{person}{Victoria Clarke}.} \bibinfo{year}{2019}\natexlab{}.
\newblock \showarticletitle{Reflecting on reflexive thematic analysis}.
\newblock \bibinfo{journal}{\emph{Qualitative research in sport, exercise and health}} \bibinfo{volume}{11}, \bibinfo{number}{4} (\bibinfo{year}{2019}), \bibinfo{pages}{589--597}.
\newblock


\bibitem[Braun and Clarke(2021)]%
        {braun2021one}
\bibfield{author}{\bibinfo{person}{Virginia Braun} {and} \bibinfo{person}{Victoria Clarke}.} \bibinfo{year}{2021}\natexlab{}.
\newblock \showarticletitle{One size fits all? What counts as quality practice in (reflexive) thematic analysis?}
\newblock \bibinfo{journal}{\emph{Qualitative research in psychology}} \bibinfo{volume}{18}, \bibinfo{number}{3} (\bibinfo{year}{2021}), \bibinfo{pages}{328--352}.
\newblock


\bibitem[Bryan-Kinns et~al\mbox{.}(2025)]%
        {bryan2025xaixarts}
\bibfield{author}{\bibinfo{person}{Nick Bryan-Kinns}, \bibinfo{person}{Shuoyang Zheng}, \bibinfo{person}{Francisco Castro}, \bibinfo{person}{Makayla Lewis}, \bibinfo{person}{Jia-Rey Chang}, \bibinfo{person}{Gabriel Vigliensoni}, \bibinfo{person}{Terence Broad}, \bibinfo{person}{Michael~Paul Clemens}, {and} \bibinfo{person}{Elizabeth Wilson}.} \bibinfo{year}{2025}\natexlab{}.
\newblock \showarticletitle{XAIxArts manifesto: explainable AI for the arts}. In \bibinfo{booktitle}{\emph{Proceedings of the Extended Abstracts of the CHI Conference on Human Factors in Computing Systems}}. \bibinfo{pages}{1--8}.
\newblock


\bibitem[Chang et~al\mbox{.}(2023)]%
        {chang_prompt_2023}
\bibfield{author}{\bibinfo{person}{Minsuk Chang}, \bibinfo{person}{Stefania Druga}, \bibinfo{person}{Alexander~J. Fiannaca}, \bibinfo{person}{Pedro Vergani}, \bibinfo{person}{Chinmay Kulkarni}, \bibinfo{person}{Carrie~J Cai}, {and} \bibinfo{person}{Michael Terry}.} \bibinfo{year}{2023}\natexlab{}.
\newblock \showarticletitle{The {Prompt} {Artists}}. In \bibinfo{booktitle}{\emph{Creativity and {Cognition}}}. \bibinfo{publisher}{ACM}, \bibinfo{address}{Virtual Event USA}, \bibinfo{pages}{75--87}.
\newblock
\showISBNx{9798400701801}
\href{https://doi.org/10.1145/3591196.3593515}{doi:\nolinkurl{10.1145/3591196.3593515}}


\bibitem[Cui et~al\mbox{.}(2025)]%
        {Cui-2025}
\bibfield{author}{\bibinfo{person}{Zhipu Cui}, \bibinfo{person}{Andong Tian}, \bibinfo{person}{Zhi Ying}, {and} \bibinfo{person}{Jialiang Lu}.} \bibinfo{year}{2025}\natexlab{}.
\newblock \showarticletitle{AC-LoRA: auto component LoRA for personalized artistic style image generation}. In \bibinfo{booktitle}{\emph{Eighth International Conference on Computer Graphics and Virtuality (ICCGV 2025)}}, \bibfield{editor}{\bibinfo{person}{Haiquan Zhao}} (Ed.). \bibinfo{publisher}{SPIE}, \bibinfo{pages}{2}.
\newblock
\href{https://doi.org/10.1117/12.3060036}{doi:\nolinkurl{10.1117/12.3060036}}


\bibitem[Donnarumma(2022)]%
        {donnarumma_against_2022}
\bibfield{author}{\bibinfo{person}{Marco Donnarumma}.} \bibinfo{year}{2022}\natexlab{}.
\newblock \showarticletitle{Against the {Norm}: {Othering} and {Otherness} in {AI} {Aesthetics}}.
\newblock \bibinfo{journal}{\emph{Digital Culture \& Society}} \bibinfo{volume}{8}, \bibinfo{number}{2} (\bibinfo{date}{Dec.} \bibinfo{year}{2022}), \bibinfo{pages}{39--66}.
\newblock
\showISSN{2364-2122, 2364-2114}
\href{https://doi.org/10.14361/dcs-2022-0205}{doi:\nolinkurl{10.14361/dcs-2022-0205}}


\bibitem[Gillespie(2024)]%
        {gillespie2024generative}
\bibfield{author}{\bibinfo{person}{Tarleton Gillespie}.} \bibinfo{year}{2024}\natexlab{}.
\newblock \showarticletitle{Generative AI and the politics of visibility}.
\newblock \bibinfo{journal}{\emph{Big Data \& Society}} \bibinfo{volume}{11}, \bibinfo{number}{2} (\bibinfo{year}{2024}), \bibinfo{pages}{20539517241252131}.
\newblock


\bibitem[Hu et~al\mbox{.}(2022)]%
        {hu2022lora}
\bibfield{author}{\bibinfo{person}{Edward~J Hu}, \bibinfo{person}{Yelong Shen}, \bibinfo{person}{Phillip Wallis}, \bibinfo{person}{Zeyuan Allen-Zhu}, \bibinfo{person}{Yuanzhi Li}, \bibinfo{person}{Shean Wang}, \bibinfo{person}{Lu Wang}, \bibinfo{person}{Weizhu Chen}, {et~al\mbox{.}}} \bibinfo{year}{2022}\natexlab{}.
\newblock \showarticletitle{Lora: Low-rank adaptation of large language models.}
\newblock \bibinfo{journal}{\emph{ICLR}} \bibinfo{volume}{1}, \bibinfo{number}{2} (\bibinfo{year}{2022}), \bibinfo{pages}{3}.
\newblock


\bibitem[Jiang et~al\mbox{.}(2023)]%
        {jiang2023ai}
\bibfield{author}{\bibinfo{person}{Harry~H Jiang}, \bibinfo{person}{Lauren Brown}, \bibinfo{person}{Jessica Cheng}, \bibinfo{person}{Mehtab Khan}, \bibinfo{person}{Abhishek Gupta}, \bibinfo{person}{Deja Workman}, \bibinfo{person}{Alex Hanna}, \bibinfo{person}{Johnathan Flowers}, {and} \bibinfo{person}{Timnit Gebru}.} \bibinfo{year}{2023}\natexlab{}.
\newblock \showarticletitle{AI Art and its Impact on Artists}. In \bibinfo{booktitle}{\emph{Proceedings of the 2023 AAAI/ACM Conference on AI, Ethics, and Society}}. \bibinfo{pages}{363--374}.
\newblock


\bibitem[Li et~al\mbox{.}(2023)]%
        {li_beyond_2023}
\bibfield{author}{\bibinfo{person}{Jingyi Li}, \bibinfo{person}{Eric Rawn}, \bibinfo{person}{Jacob Ritchie}, \bibinfo{person}{Jasper Tran~O'Leary}, {and} \bibinfo{person}{Sean Follmer}.} \bibinfo{year}{2023}\natexlab{}.
\newblock \showarticletitle{Beyond the {Artifact}: {Power} as a {Lens} for {Creativity} {Support} {Tools}}. In \bibinfo{booktitle}{\emph{Proceedings of the 36th {Annual} {ACM} {Symposium} on {User} {Interface} {Software} and {Technology}}}. \bibinfo{publisher}{ACM}, \bibinfo{address}{San Francisco CA USA}, \bibinfo{pages}{1--15}.
\newblock
\showISBNx{9798400701320}
\href{https://doi.org/10.1145/3586183.3606831}{doi:\nolinkurl{10.1145/3586183.3606831}}


\bibitem[Liao and Varshney(2021)]%
        {liao2021human}
\bibfield{author}{\bibinfo{person}{Q~Vera Liao} {and} \bibinfo{person}{Kush~R Varshney}.} \bibinfo{year}{2021}\natexlab{}.
\newblock \showarticletitle{Human-centered explainable ai (xai): From algorithms to user experiences}.
\newblock \bibinfo{journal}{\emph{arXiv preprint arXiv:2110.10790}} (\bibinfo{year}{2021}).
\newblock


\bibitem[Liao and Vaughan(2023)]%
        {liao2023ai}
\bibfield{author}{\bibinfo{person}{Q~Vera Liao} {and} \bibinfo{person}{Jennifer~Wortman Vaughan}.} \bibinfo{year}{2023}\natexlab{}.
\newblock \showarticletitle{Ai transparency in the age of llms: A human-centered research roadmap}.
\newblock \bibinfo{journal}{\emph{arXiv preprint arXiv:2306.01941}} (\bibinfo{year}{2023}).
\newblock


\bibitem[Luccioni et~al\mbox{.}(2024)]%
        {luccioni_stable_2024}
\bibfield{author}{\bibinfo{person}{Sasha Luccioni}, \bibinfo{person}{Christopher Akiki}, \bibinfo{person}{Margaret Mitchell}, {and} \bibinfo{person}{Yacine Jernite}.} \bibinfo{year}{2024}\natexlab{}.
\newblock \showarticletitle{Stable bias: {Evaluating} societal representations in diffusion models}.
\newblock \bibinfo{journal}{\emph{Advances in Neural Information Processing Systems}}  \bibinfo{volume}{36} (\bibinfo{year}{2024}).
\newblock
\urldef\tempurl%
\url{https://proceedings.neurips.cc/paper_files/paper/2023/hash/b01153e7112b347d8ed54f317840d8af-Abstract-Datasets_and_Benchmarks.html}
\showURL{%
\tempurl}


\bibitem[Lundberg(2017)]%
        {lundberg2017unified}
\bibfield{author}{\bibinfo{person}{Scott Lundberg}.} \bibinfo{year}{2017}\natexlab{}.
\newblock \showarticletitle{A unified approach to interpreting model predictions}.
\newblock \bibinfo{journal}{\emph{arXiv preprint arXiv:1705.07874}} (\bibinfo{year}{2017}).
\newblock


\bibitem[Mack et~al\mbox{.}(2024)]%
        {10.1145/3613904.3642166}
\bibfield{author}{\bibinfo{person}{Kelly~Avery Mack}, \bibinfo{person}{Rida Qadri}, \bibinfo{person}{Remi Denton}, \bibinfo{person}{Shaun~K. Kane}, {and} \bibinfo{person}{Cynthia~L. Bennett}.} \bibinfo{year}{2024}\natexlab{}.
\newblock \showarticletitle{“They only care to show us the wheelchair”: disability representation in text-to-image AI models}. In \bibinfo{booktitle}{\emph{Proceedings of the 2024 CHI Conference on Human Factors in Computing Systems}} (Honolulu, HI, USA) \emph{(\bibinfo{series}{CHI '24})}. \bibinfo{publisher}{Association for Computing Machinery}, \bibinfo{address}{New York, NY, USA}, Article \bibinfo{articleno}{288}, \bibinfo{numpages}{23}~pages.
\newblock
\showISBNx{9798400703300}
\href{https://doi.org/10.1145/3613904.3642166}{doi:\nolinkurl{10.1145/3613904.3642166}}


\bibitem[Mahdavi~Goloujeh et~al\mbox{.}(2024)]%
        {goloujeh2024promptjourney}
\bibfield{author}{\bibinfo{person}{Atefeh Mahdavi~Goloujeh}, \bibinfo{person}{Anne Sullivan}, {and} \bibinfo{person}{Brian Magerko}.} \bibinfo{year}{2024}\natexlab{}.
\newblock \showarticletitle{Is It AI or Is It Me? Understanding Users’ Prompt Journey with Text-to-Image Generative AI Tools}. In \bibinfo{booktitle}{\emph{Proceedings of the 2024 CHI Conference on Human Factors in Computing Systems}} (Honolulu, HI, USA) \emph{(\bibinfo{series}{CHI '24})}. \bibinfo{publisher}{Association for Computing Machinery}, \bibinfo{address}{New York, NY, USA}, Article \bibinfo{articleno}{183}, \bibinfo{numpages}{13}~pages.
\newblock
\showISBNx{9798400703300}
\href{https://doi.org/10.1145/3613904.3642861}{doi:\nolinkurl{10.1145/3613904.3642861}}


\bibitem[McCormack et~al\mbox{.}(2023)]%
        {mccormack_is_2023}
\bibfield{author}{\bibinfo{person}{Jon McCormack}, \bibinfo{person}{Camilo Cruz~Gambardella}, \bibinfo{person}{Nina Rajcic}, \bibinfo{person}{Stephen~James Krol}, \bibinfo{person}{Maria~Teresa Llano}, {and} \bibinfo{person}{Meng Yang}.} \bibinfo{year}{2023}\natexlab{}.
\newblock \showarticletitle{Is {Writing} {Prompts} {Really} {Making} {Art}?}
\newblock In \bibinfo{booktitle}{\emph{Artificial {Intelligence} in {Music}, {Sound}, {Art} and {Design}}}, \bibfield{editor}{\bibinfo{person}{Colin Johnson}, \bibinfo{person}{Nereida Rodríguez-Fernández}, {and} \bibinfo{person}{Sérgio~M. Rebelo}} (Eds.). Vol.~\bibinfo{volume}{13988}. \bibinfo{publisher}{Springer Nature Switzerland}, \bibinfo{address}{Cham}, \bibinfo{pages}{196--211}.
\newblock
\showISBNx{978-3-031-29955-1 978-3-031-29956-8}
\href{https://doi.org/10.1007/978-3-031-29956-8_13}{doi:\nolinkurl{10.1007/978-3-031-29956-8_13}}
\newblock
\shownote{Series Title: Lecture Notes in Computer Science}.


\bibitem[Mim et~al\mbox{.}(2024)]%
        {mim2024between}
\bibfield{author}{\bibinfo{person}{Nusrat~Jahan Mim}, \bibinfo{person}{Dipannita Nandi}, \bibinfo{person}{Sadaf~Sumyia Khan}, \bibinfo{person}{Arundhuti Dey}, {and} \bibinfo{person}{Syed~Ishtiaque Ahmed}.} \bibinfo{year}{2024}\natexlab{}.
\newblock \showarticletitle{In-between visuals and visible: The impacts of text-to-image generative ai tools on digital image-making practices in the global south}. In \bibinfo{booktitle}{\emph{Proceedings of the 2024 CHI Conference on Human Factors in Computing Systems}}. \bibinfo{pages}{1--18}.
\newblock


\bibitem[Qadri et~al\mbox{.}(2025)]%
        {qadri_ai_2025}
\bibfield{author}{\bibinfo{person}{Rida Qadri}, \bibinfo{person}{Piotr Mirowski}, {and} \bibinfo{person}{Remi Denton}.} \bibinfo{year}{2025}\natexlab{}.
\newblock \showarticletitle{{AI} and {Non}-{Western} {Art} {Worlds}: {Reimagining} {Critical} {AI} {Futures} through {Artistic} {Inquiry} and {Situated} {Dialogue}}. In \bibinfo{booktitle}{\emph{Proceedings of the 2025 {CHI} {Conference} on {Human} {Factors} in {Computing} {Systems}}} \emph{(\bibinfo{series}{{CHI} '25})}. \bibinfo{publisher}{Association for Computing Machinery}, \bibinfo{address}{New York, NY, USA}, \bibinfo{pages}{1--17}.
\newblock
\showISBNx{979-8-4007-1394-1}
\href{https://doi.org/10.1145/3706598.3714049}{doi:\nolinkurl{10.1145/3706598.3714049}}


\bibitem[Qadri et~al\mbox{.}(2023)]%
        {qadri_ais_2023}
\bibfield{author}{\bibinfo{person}{Rida Qadri}, \bibinfo{person}{Renee Shelby}, \bibinfo{person}{Cynthia~L. Bennett}, {and} \bibinfo{person}{Emily Denton}.} \bibinfo{year}{2023}\natexlab{}.
\newblock \showarticletitle{{AI}’s {Regimes} of {Representation}: {A} {Community}-centered {Study} of {Text}-to-{Image} {Models} in {South} {Asia}}. In \bibinfo{booktitle}{\emph{Proceedings of the 2023 {ACM} {Conference} on {Fairness}, {Accountability}, and {Transparency}}} \emph{(\bibinfo{series}{{FAccT} '23})}. \bibinfo{publisher}{Association for Computing Machinery}, \bibinfo{address}{New York, NY, USA}, \bibinfo{pages}{506--517}.
\newblock
\showISBNx{9798400701924}
\href{https://doi.org/10.1145/3593013.3594016}{doi:\nolinkurl{10.1145/3593013.3594016}}


\bibitem[Radford et~al\mbox{.}(2023)]%
        {radford2023robust}
\bibfield{author}{\bibinfo{person}{Alec Radford}, \bibinfo{person}{Jong~Wook Kim}, \bibinfo{person}{Tao Xu}, \bibinfo{person}{Greg Brockman}, \bibinfo{person}{Christine McLeavey}, {and} \bibinfo{person}{Ilya Sutskever}.} \bibinfo{year}{2023}\natexlab{}.
\newblock \showarticletitle{Robust speech recognition via large-scale weak supervision}. In \bibinfo{booktitle}{\emph{International conference on machine learning}}. PMLR, \bibinfo{pages}{28492--28518}.
\newblock


\bibitem[Rajcic et~al\mbox{.}(2024)]%
        {rajcic2024towards}
\bibfield{author}{\bibinfo{person}{Nina Rajcic}, \bibinfo{person}{Maria~Teresa Llano~Rodriguez}, {and} \bibinfo{person}{Jon McCormack}.} \bibinfo{year}{2024}\natexlab{}.
\newblock \showarticletitle{Towards a Diffractive Analysis of Prompt-Based Generative AI}. In \bibinfo{booktitle}{\emph{Proceedings of the CHI Conference on Human Factors in Computing Systems}}. \bibinfo{pages}{1--15}.
\newblock


\bibitem[Ribeiro et~al\mbox{.}(2016)]%
        {ribeiro2016should}
\bibfield{author}{\bibinfo{person}{Marco~Tulio Ribeiro}, \bibinfo{person}{Sameer Singh}, {and} \bibinfo{person}{Carlos Guestrin}.} \bibinfo{year}{2016}\natexlab{}.
\newblock \showarticletitle{" Why should i trust you?" Explaining the predictions of any classifier}. In \bibinfo{booktitle}{\emph{Proceedings of the 22nd ACM SIGKDD international conference on knowledge discovery and data mining}}. \bibinfo{pages}{1135--1144}.
\newblock


\bibitem[Ruiz et~al\mbox{.}(2023)]%
        {ruiz2023dreambooth}
\bibfield{author}{\bibinfo{person}{Nataniel Ruiz}, \bibinfo{person}{Yuanzhen Li}, \bibinfo{person}{Varun Jampani}, \bibinfo{person}{Yael Pritch}, \bibinfo{person}{Michael Rubinstein}, {and} \bibinfo{person}{Kfir Aberman}.} \bibinfo{year}{2023}\natexlab{}.
\newblock \showarticletitle{Dreambooth: Fine tuning text-to-image diffusion models for subject-driven generation}. In \bibinfo{booktitle}{\emph{Proceedings of the IEEE/CVF conference on computer vision and pattern recognition}}. \bibinfo{pages}{22500--22510}.
\newblock


\bibitem[Sanchez(2023)]%
        {sanchez_examining_2023}
\bibfield{author}{\bibinfo{person}{Téo Sanchez}.} \bibinfo{year}{2023}\natexlab{}.
\newblock \showarticletitle{Examining the {Text}-to-{Image} {Community} of {Practice}: {Why} and {How} do {People} {Prompt} {Generative} {AIs}?}. In \bibinfo{booktitle}{\emph{Creativity and {Cognition}}}. \bibinfo{publisher}{ACM}, \bibinfo{address}{Virtual Event USA}, \bibinfo{pages}{43--61}.
\newblock
\showISBNx{9798400701801}
\href{https://doi.org/10.1145/3591196.3593051}{doi:\nolinkurl{10.1145/3591196.3593051}}


\bibitem[Shelby et~al\mbox{.}(2024)]%
        {shelby2024generative}
\bibfield{author}{\bibinfo{person}{Renee Shelby}, \bibinfo{person}{Shalaleh Rismani}, {and} \bibinfo{person}{Negar Rostamzadeh}.} \bibinfo{year}{2024}\natexlab{}.
\newblock \showarticletitle{Generative AI in Creative Practice: ML-Artist Folk Theories of T2I Use, Harm, and Harm-Reduction}. In \bibinfo{booktitle}{\emph{Proceedings of the CHI Conference on Human Factors in Computing Systems}}. \bibinfo{pages}{1--17}.
\newblock


\bibitem[Shneiderman(2007)]%
        {shneiderman2007creativity}
\bibfield{author}{\bibinfo{person}{Ben Shneiderman}.} \bibinfo{year}{2007}\natexlab{}.
\newblock \showarticletitle{Creativity support tools: accelerating discovery and innovation}.
\newblock \bibinfo{journal}{\emph{Commun. ACM}} \bibinfo{volume}{50}, \bibinfo{number}{12} (\bibinfo{year}{2007}), \bibinfo{pages}{20--32}.
\newblock


\bibitem[Smith et~al\mbox{.}(2026)]%
        {smith2026noise}
\bibfield{author}{\bibinfo{person}{James Smith}, \bibinfo{person}{Shm~Garanganao Almeda}, \bibinfo{person}{Timothy~J Aveni}, \bibinfo{person}{Anya Agarwal}, {and} \bibinfo{person}{Bjoern Hartmann}.} \bibinfo{year}{2026}\natexlab{}.
\newblock \showarticletitle{Noise Pilot: Enabling Artistic Workflow Composition with Diffusion-Based Image Generation}. In \bibinfo{booktitle}{\emph{Proceedings of the 2026 CHI Conference on Human Factors in Computing Systems}}. \bibinfo{pages}{1--20}.
\newblock


\bibitem[Sundararajan et~al\mbox{.}(2017)]%
        {sundararajan2017axiomatic}
\bibfield{author}{\bibinfo{person}{Mukund Sundararajan}, \bibinfo{person}{Ankur Taly}, {and} \bibinfo{person}{Qiqi Yan}.} \bibinfo{year}{2017}\natexlab{}.
\newblock \showarticletitle{Axiomatic attribution for deep networks}. In \bibinfo{booktitle}{\emph{International conference on machine learning}}. PMLR, \bibinfo{pages}{3319--3328}.
\newblock


\bibitem[Taylor et~al\mbox{.}(2025)]%
        {taylor_-straightening_2025}
\bibfield{author}{\bibinfo{person}{Jordan Taylor}, \bibinfo{person}{Joel Mire}, \bibinfo{person}{Franchesca Spektor}, \bibinfo{person}{Alicia DeVrio}, \bibinfo{person}{Maarten Sap}, \bibinfo{person}{Haiyi Zhu}, {and} \bibinfo{person}{Sarah~E Fox}.} \bibinfo{year}{2025}\natexlab{}.
\newblock \showarticletitle{Un-{Straightening} {Generative} {AI}: {How} {Queer} {Artists} {Surface} and {Challenge} {Model} {Normativity}}. In \bibinfo{booktitle}{\emph{Proceedings of the 2025 {ACM} {Conference} on {Fairness}, {Accountability}, and {Transparency}}}. \bibinfo{publisher}{ACM}, \bibinfo{address}{Athens Greece}, \bibinfo{pages}{951--963}.
\newblock
\href{https://doi.org/10.1145/3715275.3732061}{doi:\nolinkurl{10.1145/3715275.3732061}}


\end{thebibliography}

\appendix
\section{Story Task Article}
\label{app:story}
\textbf{93-year-old grandmother and grandson complete goal of visiting all 63 U.S. national parks}

By Caitlin O’Kane (CBS News)

By the time she was 85 years old, Joy Ryan of Duncan Falls, Ohio, had never seen the ocean or mountains. Now, she's 93 years old and has seen every corner of the U.S. – after visiting all 63 U.S. National Parks. Joy went on the epic journey with her grandson, Brad Ryan, now age 42, who was first inspired to travel with is grandmother in 2015. 

"When I learned she had never seen the great wildernesses of America – deserts, mountains, oceans, you name it – I thought that was something that would haunt me if I didn't intervene in some way," Brad told CBS News in October 2022.

So, Brad decided to take her to the Great Smoky Mountains National Park, and that trip sparked an idea. "The more I kept reading about the parks and I saw how close they were to one another, that we could make this giant loop, it became an obsession," 

Brad said. "And all I could think about was, 'I've got to see Grandma Joy at Old Faithful, I've got to see her at the Redwoods, I've got to see her at the Grand Canyon. I just have to do this, I just have to have those memories for my own long-term happiness. I think we're two peas in a pod when it comes to just our desire for travel, adventure, connection.’”
 The pair started planning road trips, hitting multiple parks each time they embarked on the road. While the pair have found themselves on many adventures during their travels – whale watching in the Channel Islands and seeing larger-than-life trees in the Redwood Forest – some of their best moments happen in the car.
 
"We hit the road together and we start talking about our lives," Brad said. "And she told me things in her 80s and 90s about her life, some of the difficult things that she's been through, that she's never spoken to anybody in here life. And I was able to open up to her about some of the trials and tribulations of my own life. That's what, I think, is so powerful about the open road, is that you only can drive so far before those memories start to creep forward."

\end{document}